# DAΦNE OPERATION WITH THE FINUDA EXPERIMENT

C. Milardi, D. Alesini, G. Benedetti, M.E. Biagini, C. Biscari, R. Boni, M. Boscolo, A. Clozza,
G. Delle Monache, G. Di Pirro, A. Drago, A. Gallo, A. Ghigo, S. Guiducci, M. Incurvati, C. Ligi,
F. Marcellini, G. Mazzitelli, L. Pellegrino, M.A. Preger, P. Raimondi, R. Ricci, U. Rotundo,
C. Sanelli, M. Serio, F. Sgamma, B. Spataro, A. Stecchi, A. Stella, C. Vaccarezza, M. Vescovi,
M. Zobov, LNF-INFN, Frascati, Italy

*Abstract*

DAΦNE operation restarted in September 2003, after a six months shut-down for the installation of FINUDA, a magnetic detector dedicated to the study of hypernuclear physics. FINUDA is the third experiment running on DAΦNE and operates while keeping on place the other detector KLOE. During the shut-down both Interaction Regions have been equipped with remotely controlled rotating quadrupoles in order to operate at different solenoid fields. Among many other hardware upgrades one of the most significant is the reshaping of the wiggler pole profile to improve the field quality and the machine dynamic aperture. Commissioning of the collider in the new configuration has been completed in short time. The peak luminosity delivered to FINUDA has reached $6 \cdot 10^{31}$ s$^{-1}$ cm$^{-2}$, with a daily integrated value close to 4 pb$^{-1}$.

## INTRODUCTION

DAΦNE is the Frascati electron-positron collider running at the energy of the Φ resonance [1]. It is based on two independent rings merging in two straight sections the Interaction Regions (IR) and on an injection system including a Linac, an intermediate damping ring and 180 meter long transfer lines joining the different accelerators.

The KLOE detector is permanently installed in IR1, while the DEAR and FINUDA experiments can be placed, one at a time, in IR2. The KLOE experiment studies CP violation in kaon decays and DEAR investigates exotic atoms.

The FINUDA experiment [2] consists in a magnetic spectrometer designed for the study of hypernuclear spectroscopy. With its cylindrical geometry around the collider interaction point (IP), it represents a relevant novelty element with respect to the existing experiments relying on fixed target setups. In fact, being a 2π detector, FINUDA can observe both the particles emitted in the hypernucleus formation and in its decay and provides a larger acceptance since the kaons are emitted isotropically. Moreover, the low energy DAΦNE kaons can be stopped in thinner targets, thus improving the resolution in the hypernuclear level measurements.

## LAYOUT EVOLUTION

A top luminosity of ≈ $7.5 \cdot 10^{31}$ cm$^{-2}$ s$^{-1}$ was reached during the year 2002. To go further towards luminosities beyond of $10^{32}$ cm$^{-2}$ s$^{-1}$ it was necessary to undertake major upgrades in the collider. The magnetic layout of the DAΦNE rings has been deeply reorganized, new IRs and the FINUDA detector have been installed. The straight sections used for injection have been disassembled and completely rearranged by adding a new sextupole and two quadrupoles and removing one out of the 3 injection kickers, which was not used. All the wigglers have been modified by changing their pole profile.

### Interaction Regions

The KLOE IR has been rebuilt relying on the experience gathered during the runs for the DEAR experiment, where it was possible to let 100 consecutive bunches collide, while keeping the $\beta_y^*$ low at the hourglass limit. The permanent magnet quadrupole triplets have been substituted with doublet ones; the inner quadrupoles, the closest to the interaction point (IP), have been removed and the outer ones have been strengthened, going from a FDF configuration to a DF one. All the quadrupoles have been equipped with independently rotating supports in order to improve the coupling correction efficiency and the ring flexibility. In this way it is possible to set the quadrupole rotation for arbitrary values of the KLOE solenoidal field. Printed circuit quadrupoles have been added around the IP for diagnostic purposes. The thin vacuum chamber has been substituted with a new one in beryllium alloy (ALBeMet) suitable for the new element configuration.

The FINUDA IR relays on four permanent magnet quadrupoles placed inside the FINUDA 1.1 T solenoidal field and on four conventional quadrupoles installed outside it. The new feature allowing quadrupole rotation has been integrated in the FINUDA IR original design; all the permanent quadrupoles can rotate independently within a range of 135 degrees and the conventional ones within 23 degrees. This mechanical solution allows for a wide operation flexibility, ranging from the low-β configuration suitable for collisions to the high-β one required for an efficient beam separation. Operation with the solenoid off is also possible.

### Wiggler upgrade

Beam measurements performed during 2002 pointed out relevant nonlinear terms in the wigglers. Simulations confirmed that such terms reduce the dynamic aperture and the beam lifetime, affecting beam-beam behaviour and beam dynamics. All these features led to a significant luminosity limitation.

The wiggler magnets have been upgraded [3], in order to improve the field quality; each pole profile has been modified by adding longitudinally and horizontally shimmed plates on the poles. Moreover an extra sextupole component has been added on one of the terminal poles in each wiggler.

The upgraded wigglers showed a significant reduction of the $2^{nd}$ and $3^{rd}$ order terms in the field, around the wiggling trajectory, the latter being responsible for the quadratic behaviour of the horizontal betatron tune versus beam displacement in the wiggler.

Tests with the beams confirmed the reduction of the non-linear terms, as well as an improvement by a factor 2 in the dynamic aperture and in the energy acceptance (see Fig. 1).

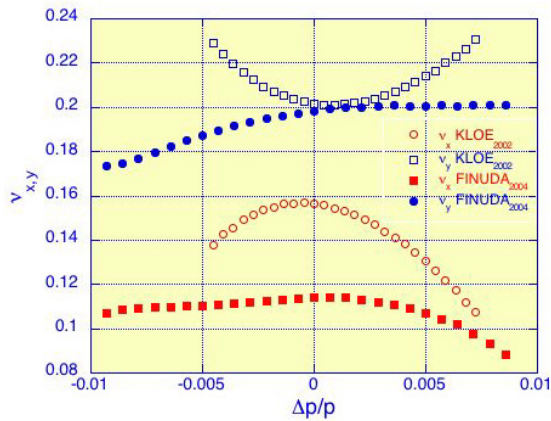

Figure 1: Tune shift dependence on energy measured in the positron ring before and after the wiggler upgrade.

## HARDWARE MAINTENANCE AND UPGRADE

During the shut-down all the collider hardware components have been carefully checked and maintained, substituted and upgraded where necessary.

Half of the ion clearing electrodes, which cope with ion induced instability in the electron ring, have been substituted with welding-free devices.

In DAΦNE tungsten scrapers are used to reduce background showers on the detectors. The scrapers are tapered by means of thin copper shields to avoid detrimental effects on the overall impedance. These shields have been modified because they were intercepting the beam. The tapers on the horizontal scrapers have been removed while the vertical ones have been properly adjusted. Some damaged bellows have been substituted and their structure reinforced by means of pin insertion.

The strip-lines used to measure the beam position along the Transfer Lines have been equipped with new electronics in order to improve the trajectory control and to speed up the injection process. For the same reason the Linac 50 Hz operation mode has been implemented: it is now possible to inject both beams at 50 Hz in the damping ring and at 2 Hz in the Main Rings.

## TUNING THE NEW DAΦNE CONFIGURATION

### Model

All the modifications in the DAΦNE layout have been included in the machine model. Special care has been devoted to the new wiggler model [4] which has been described as a sequence of 2 m long hard edge dipoles and straight sections representing the linear part of the beam motion. The non-linear terms have been introduced by adding at both pole sides higher order multipoles. All the physical parameters have been derived from the vertical component of the magnetic field in the horizontal midplane, measured on a spare modified wiggler.

It is worth reminding that the new KLOE IR gives a smaller contribution to the main rings natural chromaticity, which, due to the focusing sextupole in the wigglers, is almost zero in the horizontal plane.

The new DAΦNE model [5] has been validated by comparing its predictions with the beam measurements. Simulations and experimental data are in satisfactory agreement both for the linear part (betatron function, betatron tunes and dispersion function) and for the non-linear one (second order dispersion, second order chromaticity and tune shift on amplitude).

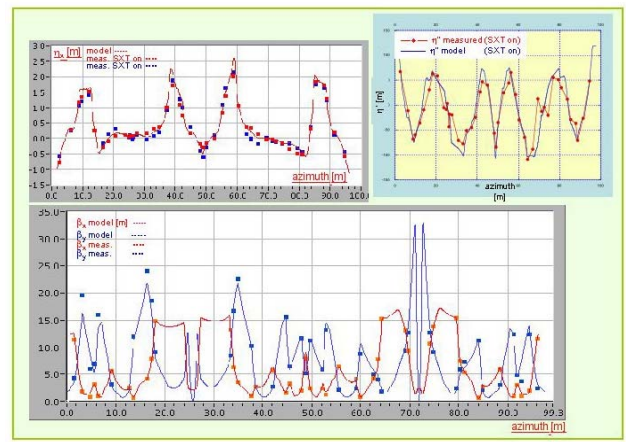

Figure 2: DAΦNE model compared with beam measurements – up: first order and second order dispersion; down: beta functions.

### DAΦNE Optics for collision at FINUDA IP

To perform collisions at the FINUDA IP, the KLOE solenoidal field (Bdl = 2.4 Tm) is off since it represents a huge perturbation with respect to the ring magnetic structure (Bρ = 1.7 Tm).

A new optics for this operation mode has been designed. The horizontal beam emittance has been reduced by 44%, becoming $\varepsilon_x$ = .34 μm, allowing for a separation of ~ 13 $\sigma_x$ at the first parasitic crossing, 0.4 m from the IP. In fact, in order to achieve collisions with 100 consecutive bunches, it is mandatory to reduce the parasitic crossings.

The dynamic aperture has been improved by optimizing the relative phase advance between sextupoles in their new configuration. The betatron functions at the IP have been set to $\beta_x^* = 2.33$ m, a reasonable trade off between the need to keep $\beta_x$ low at the IP and at the first parasitic crossing, and $\beta_y^* = .024$ m, compatible with the limit imposed by the hourglass effect. The horizontal crossing angle at the IP has been set to $\theta_x = .021$ rad in order to reduce background on the detector.

With these parameters the Piwinski angle is $\phi = .29$ similar to the values already used in the past.

*Coupling correction*

A local betatron coupling correction has been implemented. The coupling terms from the measured Response Matrix for the two rings (C matrix) have been minimized by means of the 8 permanent quadrupoles rotations in the two IRs ($\Delta\phi$ array), using the M matrix computed from the machine model and describing the coupling terms as a function of the IR quadrupole rotation [6]. The linear equation system:

$$M\Delta\phi = C$$

has been solved by using the singular value decomposition method, and, after few iterations, the rms value of the coupling terms has been reduced by 40%.

The fine betatron coupling correction has been obtained, as usual, using skew quadrupoles achieving a final coupling $\kappa = .3\%$.

## HIGH CURRENT ISSUES

The DAFNE rings are equipped with transverse and longitudinal feedbacks to cope with the effects of coupled-bunch instabilities. After the DAΦNE upgrade all the feedbacks have been carefully tuned, especially the horizontal one whose power amplifiers have been improved [7].

Despite the relevant evolution, the ring impedance, obtained from the bunch length measurements has turned out to be almost unchanged (see Fig. 3) [8], yielding a reasonable indication for an unaffected beam instability scenario.

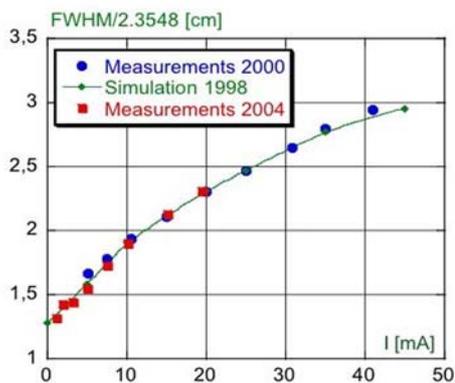

Figure 3: Bunch length measurements and theoretical predictions as a function of the bunch current for the positron ring.

Hundred bunches per beam have been routinely put in collision, with maximum colliding currents: $I^+ = .8$ A, $I^- = 1.1$ A. A positron current threshold has been observed at start-up and cured by a careful feedback systems tuning.

## LUMINOSITY

In a short time DAΦNE luminosity has reached performances comparable to the best obtained in 2002, delivering to FINUDA a peak luminosity, $L_{peak} = 0.6 \cdot 10^{32}$ cm$^{-2}$s$^{-1}$ and a total integrated luminosity close to $L_{int} = 256$ pb$^{-1}$ with excellent background rates.

Peak luminosity was mainly limited by the residual parasitic crossing in the KLOE IR, where the two beams cannot be efficiently separated in the vertical plane due to the presence of the permanent magnet quads.

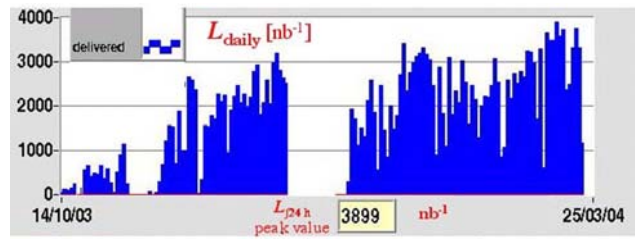

Figure 4: Daily integrated luminosity during the FINUDA runs.

## CONCLUSIONS

The DAΦNE collider has been successfully commissioned after a major upgrade. During this period all the efforts have been addressed to check the effectiveness of the changes, and to improve the collider performances. The FINUDA experiment has completed the first stage of its scientific program and data analysis is under way.

Presently DAΦNE has restarted operation for the KLOE experiment and has reached a peak luminosity $L_{peck} = 0.85 \cdot 10^{32}$ cm$^{-2}$s$^{-1}$ and currents exceeding 1.1 A in both beams have been stored in collision.